\documentclass[final,3p,times,twocolumn]{elsarticle}
\usepackage{amssymb}
\usepackage[dvipsnames]{xcolor}
\usepackage{amsmath}
\usepackage{soul,hyperref}
\journal{Physics Letters B}

\begin{document}

\begin{frontmatter}

\title{Off-shell generalized parton distributions and form factors of the pion}

\author[1,2]{Wojciech Broniowski}
\ead{Wojciech.Broniowski@ifj.edu.pl}

\author[2]{Vanamali Shastry}
\ead{vanamalishastry@gmail.com}

\author[3]{Enrique Ruiz Arriola}
\ead{earriola@ugr.es}

\affiliation[1]{organization={H. Niewodnicza\'nski Institute of Nuclear Physics PAN},
postcode={31-342},
city={Cracow},
country={Poland}}

\affiliation[2]{organization={Institute of Physics},
addressline={Jan Kochanowski University},
postcode={25-406},
city={Kielce},
country={Poland}}

\affiliation[3]{organization={Departamento de F\'{\i}sica At\'{\o}mica, Molecular y Nuclear and Instituto Carlos I de  F{\'\i}sica Te\'{o}rica y Computacional},
addressline={Universidad de Granada},
postcode={E-18071},
city={Granada},
country={Spain}}

\begin{abstract}
Off-shell effects in generalized parton distributions (GPDs) of the
pion, appearing, e.g., in the Sullivan process, are considered. Due to
the lack of crossing symmetry, the moments of GPDs involve also odd
powers of the skewness (longitudinal momentum transfer) parameter,
which results in emergence of new off-shell form factors. With
current-algebra techniques, we derive exact relations between the four
off-shell gravitational form factors of the pion, in analogy to the
electromagnetic case. Our results place stringent constraints on the
off-shell GPDs of the pion. We provide an explicit realization in
terms of a chiral quark model, where we show that the off-shell
effects in GPDs are potentially significant in modeling physical
processes and should not be neglected.
\end{abstract}

\begin{keyword}


generalized parton distributions \sep off-shell pions \sep Sullivan process

\end{keyword}

\end{frontmatter}


\bibliographystyle{elsarticle-num-names} 

\begin{figure}[t]
\centering
\includegraphics[angle=0,width=0.36 \textwidth]{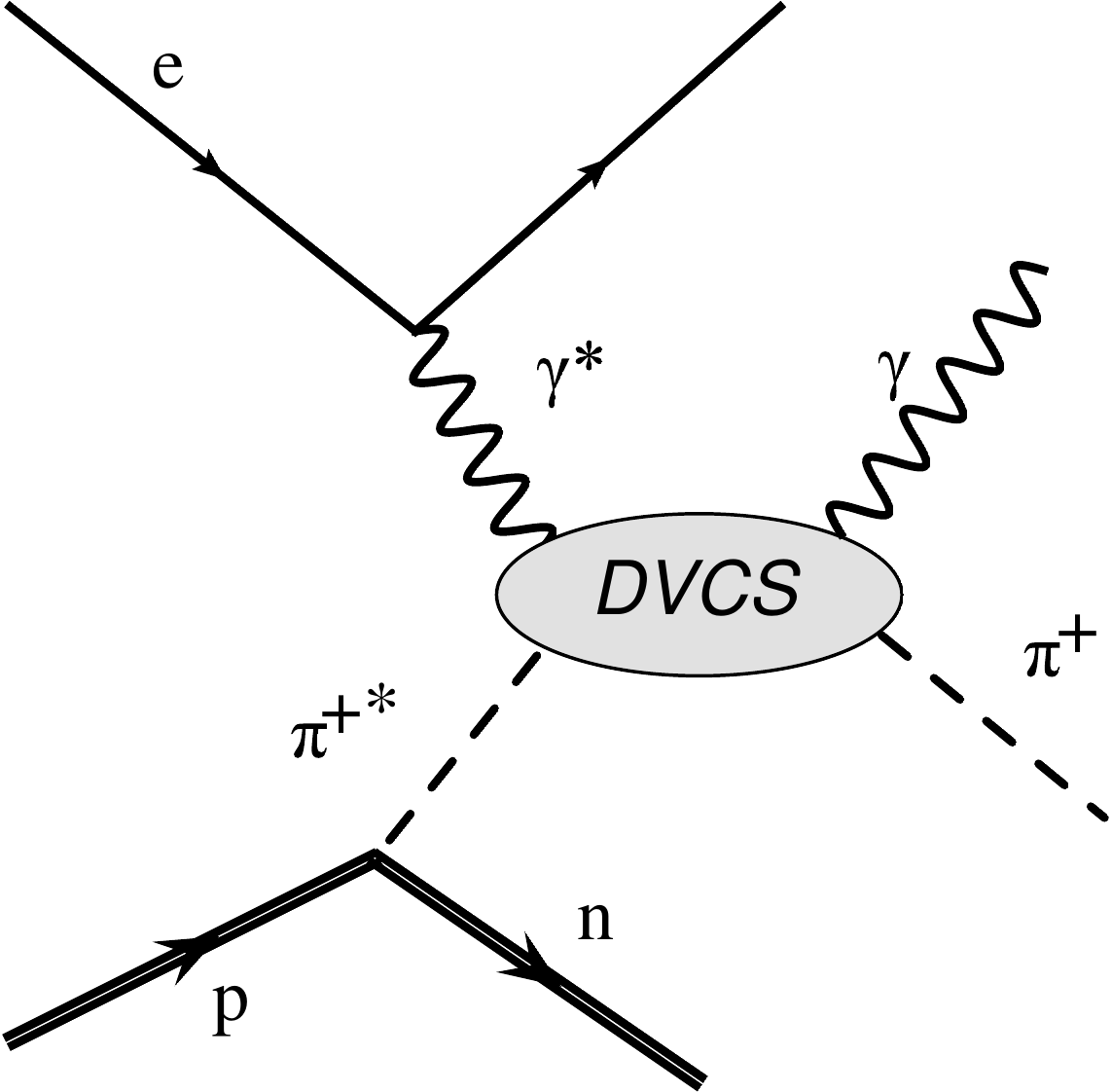} 
\vspace{-1mm}
\caption{Sullivan process for the pion electroproduction off the proton, containing the deeply virtual Compton scattering (DVCS) amplitude involving GPDs. 
Asterisks indicate off-shellness.
\label{fig:sul}}
\end{figure}

In recent papers~\cite{Aguilar:2019teb,Chavez:2021koz}, accessibility
of the pion GPDs~\cite{Ji:1996nm,Radyushkin:1997ki} via the Sullivan
process~\cite{PhysRevD.5.1732} in future electron-ion colliders has
been studied, with the conclusion that it may soon fall within
experimental reach. Since the corresponding amplitude involves an
off-shell pion (cf. Fig.~\ref{fig:sul}), one needs to care about the
possible off-shellness issues in such processes and in the GPDs
themselves. Whereas these admittedly unmeasurable effects would cancel in an ultimate complete 
calculation of $e^+ p \to e^+ \pi^+n \gamma$,\footnote{Such a calculation, involving all possible QCD diagrams, is not conceivable in a phenomenological approach, but could be hoped for in lattice simulations in distant future.} they {\em unavoidably} do show up in
phenomenological approaches which treat the building blocks $p \to \pi^{+\ast} n$ and $\gamma^\ast \pi^{+\ast} \to \gamma \pi^+$ as
independent subprocesses.

The pion, being a pseudo-Goldstone boson of the spontaneously broken
chiral symmetry, is by far the simplest hadron. Yet, its
nonperturbative structure is rich, as can be revealed with the methods
involving GPDs (the 3D hadronic tomography~\cite{Burkardt:2000za}). In
this Letter we show that relations between various off-shell form factors of the
pion (electromagnetic, gravitational) provide highly non-trivial
constraints for the structure of the off-shell GPDs. These relations, which we
extend to the case of the four off-shell gravitational form factors,
follow from the Ward-Takahashi identities (WTI).

The off-shell quark and gluon GPDs of the pion are defined, in the
notation of~\cite{Diehl:2003ny}, via the matrix element of bilocal
fields
\begin{eqnarray}
&&\hspace{-7mm} \delta_{ab}\delta_{\alpha\beta}{H}^{0}(x,\xi,t,p_i^2,p_f^2)+i\epsilon^{abc}\tau^c_{\alpha\beta}{H}^{1}(x,\xi,t,p_i^2,p_f^2) = \nonumber \\
&&\hspace{-5mm} \left .  \int \! \frac{d z^-}{4\pi} e^{i x \, P^+ z^-} \langle\pi^b(p_f)|\overline{\psi}_\alpha(-\tfrac{z}{2}) \gamma^+ \psi_\beta(\tfrac{z}{2})|\pi^a(p_i)\rangle \right |_{\substack{z^+=0\\z^\perp=0}}, \nonumber \\ 
&& \hspace{-7mm}  \delta_{ab}{H}^{g}(x,\xi,t,p_i^2,p_f^2)= \label{eq:Hdef}  \\
&& \hspace{-5mm} \left . \int \! \frac{d z^-}{2\pi P^+} e^{i x \, P^+ z^-} \langle\pi^b(p_f)| F^{+\mu}(-\tfrac{z}{2}) {F_\mu}^+(\tfrac{z}{2})  |\pi^a(p_i)\rangle \right |_{\substack{z^+=0\\z^\perp=0}}, \nonumber
\end{eqnarray}
where $\psi$ indicates the quark field, $F^{\mu \nu}$ is the gluon
field tensor, $a$, $b$, and $c$ are the isospin indices of the pion,
$\alpha$ and $\beta$ are the quark flavors, and summation over color
is implicit.  The subscripts $0,1$ denote the isospin of the quark
GPDs. The light-cone indices are $v^\pm= (v^0 \pm v^3)/\sqrt{2}$. In
the assumed light-cone gauge the link operators do not appear. The
adopted symmetric notation for the kinematic variables is
\begin{eqnarray}
P^\mu=\tfrac{1}{2} (p_f^\mu + p_i^\mu), \;\; q^\mu=p_f^\mu - p_i^\mu,  \;\; \xi=-\frac{q^+}{2P^+}, \;\;  t=q^2. \nonumber 
~\hspace{-7mm} \\
\label{eq:not}
\end{eqnarray}
In the partonic interpretation (in the on-shell case)
$p_f^2=p_i^2=m_\pi^2$, while $(x+\xi)P^+$ is the longitudinal momentum
carried by the struck parton. The GPDs $H^0, H^1$ and $H^g$ are scale
dependent objects which follow the DGLAP-ERBL QCD evolution
equations~\cite{Muller:1994ses,Diehl:2003ny}.  We note that
off-shellness of the initial and final hadronic states (pions) does
not affect the QCD evolution kernel in the assumed Bjorken limit.

For $p_f^2=p_i^2 $ the crossing symmetry (time-reversal)
makes the amplitudes in Eq.~(\ref{eq:Hdef}) even functions of the
skewness parameter $\xi$.  This feature no longer holds with general
off-shellness, when $p_f^2 \neq p_i^2$, as happens in the Sullivan process of Fig.~\ref{fig:sul}.  In particular, in that case
the $x$-moments of the GPDs involve also odd powers of $\xi$,
\begin{eqnarray}
&& \hspace{-7mm} \int_{-1}^1 dx \, x^j H^s(x,\xi,t,p_i^2,p_f^2)=\sum_{i=0}^{j+1} A^{s}_{j,i}(t,p_i^2,p_f^2) \xi^i,  \label{eq:poly}
\end{eqnarray}
where $s=0,1,g$.  Among the generalized (off-shell) form factors
$A^{s}_{j,i}$, those related
to the conserved electromagnetic and energy-stress tensor currents (the two lowest $x$-moments)
are particularly important. They do not depend on the factorization
scale, and therefore are independent of the QCD
evolution. Explicitly, 
\begin{eqnarray}
&& \hspace{-7mm}  \int_{-1}^1 dx \,  {H}^{1}=2(F - G\xi),  \label{eq:poly2} \\
&& \hspace{-7mm}  \int_{-1}^1 dx \,  x[H^{0}+H^{g}]=\theta_2 - \theta_3\xi - \theta_1\xi^2, \nonumber
\end{eqnarray}
where the form factors are functions of $(t,p_i^2,p_f^2)$. Thus the above conditions depend nontrivially on $t$ and the
off-shellness parameters.

From a dynamical point of view, the complementary role of the
electromagnetic and gravitational form factors at zero momentum
transfer, ensuring a proper normalization of the Bethe-Salpeter bound
state equation for on-shell states, was recognized long
ago~\cite{Nishijima:1967byg}. Here we consider a general
off-shell case. First, we
recall for completeness the results obtained for the off-shell effects in
the pion charge form factors~\cite{Naus:1989em,Rudy:1994qb}.  The
(one-particle irreducible, renormalized) pion-photon vertex (we take
the positively charged pion for definiteness) has the general covariant structure
\begin{eqnarray}
\Gamma^\mu(p_i,p_f)=2P^\mu F(t,p_i^2,p_f^2)+q^\mu G(t,p_i^2,p_f^2). \label{eq:Vdec}
\end{eqnarray}
Next, one considers the WTI for the full $\pi\pi\gamma$ vertex (with
unamputated external pion propagators):
\begin{eqnarray}
&& \hspace{-7mm} (2\pi)^4 \delta^{(4)}(p_f-p_i-q) G^{\mu}(p_i,p_f) = \int d^4x\,d^4y\,d^4z  \nonumber \\ 
&& \times \, e^{i(p_f\cdot x - p_i\cdot y - q \cdot z)} \langle 0 | T\left( \phi^+(x) \phi^-(y) J^{\mu}(z) \right) | 0 \rangle,
\end{eqnarray}
where the standard use of the covariantized time order product (or the $T^\ast$
product), with the time derivatives pulled outside, is understood from
now on.  Using the current algebra of the vector and axial
currents~\cite{Schnitzer:1967zzb}, one finds
\begin{eqnarray}
&& \hspace{-7mm}  (2\pi)^4 \delta^{(4)}(p_f-p_i-q) q_\mu G^{\mu}(p_i,p_f) = i \int d^4x\,d^4y  \nonumber \\ 
&& \times \,\left ( e^{i p_f \cdot (x - y)} - e^{i p_i \cdot (x - y)} \right ) \langle 0 | T \left( \phi^+(x) \phi^-(y) \right) | 0 \rangle, \label{eq:s3}
\end{eqnarray}
which yields 
$q_\mu G^{\mu}(p_i,p_f) = \Delta(p_f^2) - \Delta(p_i^2)$,  
where $\Delta(p^2)$ denotes the pion propagator. 
Next, one passes to the irreducible vertex $\Gamma^\mu$ by the standard leg amputation procedure, namely 
\begin{eqnarray}
\Gamma^{\mu}(p_i,p_f) = \left(i \Delta(p_f^2)\right)^{-1}G^{\mu}(p_i,p_f)\left(i \Delta(p_i^2)\right)^{-1}, \label{eq:leg}
\end{eqnarray}
which finally gives the WTI of \cite{Rudy:1994qb},
\begin{eqnarray}
q_\mu \Gamma^{\mu}(p_i,p_f) = \Delta\!^{-1}(p_f^2) - \Delta\!^{-1}(p_i^2), \label{eq:irr}
\end{eqnarray} 
with $\Delta(p^2)$ denoting the pion propagator. 
From Eq.~(\ref{eq:Vdec}) one finds that $q_\mu\Gamma^\mu=(p_f^2\!-\!p_i^2) F + t G$, hence the relation
\begin{eqnarray}
&& \hspace{-7mm} (p_f^2\!-\!p_i^2) F(t,p_i^2,p_f^2) + t G(t,p_i^2,p_f^2) =  \Delta\!^{-1}(p_f^2)- \Delta\!^{-1}(p_i^2) \nonumber \\
&& ~ \hspace{-10mm}  \label{eq:rule}
\end{eqnarray}
follows.
At $t=0$ (under the natural assumption that $G(t,p_i^2,p_f^2)$ is not singular) one obtains
\begin{eqnarray}
\Delta\!^{-1}(p_f^2)- \Delta\!^{-1}(p_i^2) = (p_f^2\!-\!p_i^2) F(0,p_i^2,p_f^2), \label{eq:rulet0}
\end{eqnarray} 
therefore 
\begin{eqnarray}
G(t,p_i^2,p_f^2) = \frac{(p_f^2\!-\!p_i^2)}{t} \left [  F(0,p_i^2,p_f^2) -  F(t,p_i^2,p_f^2)\right ] \label{eq:Grel}
\end{eqnarray}
and $ G(0,p_i^2,p_f^2) =(p_i^2\!-\!p_f^2)dF(t,p_i^2,p_f^2)/dt |_{t=0}$.
The simplicity of the result should not cover up its depth, namely, the off-shell $G$ form factor is completely expressible via the off-shell $F$ form factor.
Further, at the pion pole $\Delta\!^{-1}(m_\pi^2)=0$, hence one finds from Eq.~(\ref{eq:rulet0}) that the half-off shell form factors at $t=0$ are
\begin{eqnarray}
F(0,m_\pi^2,p^2) = F(0,p^2,m_\pi^2) = \frac{\Delta\!^{-1}(p^2)}{(p^2-m_\pi^2)}. \label{eq:halfo}
\end{eqnarray}
Taking the limit $p^2 \to m_\pi^2$ one gets $F(0,m_\pi^2,m_\pi^2)=1$, which is the charge normalization of the pion. 
For equal off-shellness of the initial and final pion, Eq.~(\ref{eq:Grel}) yields immediately $G(t,p^2,p^2)=0$, which is a manifestation of the crossing symmetry.
The form factor $G(t,p^2,m_\pi^2)/p^2$ has been recently studied phenomenologically in a quark model in~\cite{Choi:2019nvk}.

Now we pass to  novel results for the off-shell gravitational form factors.  
The gravitational vertex has the general tensorial structure 
\begin{eqnarray}
&& \hspace{-7mm} \Gamma^{\mu \nu} =  \tfrac{1}{2}  [(q^2 g^{\mu \nu} - q^\mu q^\nu) \theta_1+  4 P^\mu P^\nu \theta_2   \nonumber \\ 
&& + 2 (q^\mu P^\nu + q^\nu P_\mu) \theta_3 - g^{\mu \nu} \, \theta_4], \label{eq:the}
\end{eqnarray}
The form factors $\theta_1$ and $\theta_2$ are even under the crossing symmetry, whereas  $\theta_3$ and $\theta_4$ are odd.
The WTI for the gravitational vertex can be derived as follows: the full vertex is defined as
\begin{eqnarray}
&& \hspace{-7mm} (2\pi)^4 \delta^{(4)}(p_f-p_i-q) G^{\mu\nu}(p_i,p_f) = \int d^4x\,d^4y\,d^4z  \nonumber \\ 
&& \times \, e^{i(p_f\cdot x - p_i\cdot y - q \cdot z)} \langle 0 | T \left( \phi^+(x) \phi^-(y) \Theta^{\mu\nu}(z) \right) | 0 \rangle,
\end{eqnarray}
where $\Theta^{\mu\nu}$ is the energy-stress tensor (involving quarks and gluons), obtained by differentiating the action with respect to the metric tensor. It is conserved, $\partial_\mu \Theta^{\mu\nu}=0$. Current algebra yields the relation (holding for the PCAC pion, not necessary an elementary field)~\cite{Raman:1971jg}
\begin{eqnarray}
&& \hspace{-7mm} (2\pi)^4 \delta^{(4)}(p_f-p_i-q) q_\mu G^{\mu \nu}(p_i,p_f) = i \int d^4x\,d^4y \times \nonumber \\ 
&& \left ( p_i^\nu e^{i p_i \cdot (x - y)} - p_f^\nu e^{i p_f \cdot (x - y)} \right ) \langle 0 | T \left( \phi^+(x) \phi^-(y) \right) | 0 \rangle, \nonumber \\ \label{eq:s4}
\end{eqnarray}
hence
$q_\mu G^{\mu \nu}(p_i,p_f) = p_i^\nu  \Delta(p_i^2) - p_f^\nu \Delta(p_f^2)$. 
Finally, the irreducible gravitational vertex becomes
\begin{eqnarray}
&& \hspace{-7mm} q_\mu \Gamma^{\mu \nu}(p_i,p_f) = p_i^\nu  \Delta\!^{-1}(p_f^2) -  p_f^\nu \Delta\!^{-1}(p_f^2) = \label{eq:WT2} \\
&&  P^\nu [ \Delta\!^{-1}(p_f^2) - \Delta\!^{-1}(p_i^2)] - \tfrac{1}{2} q^\nu [ \Delta\!^{-1}(p_f^2) + \Delta\!^{-1}(p_i^2)]. \nonumber
\end{eqnarray}
Remarkably, this relation was first obtained by Brout and Englert~\cite{Brout:1966oea} using just the general gravitational covariance.
On the other hand, from Eq.~(\ref{eq:the}) we get
\begin{eqnarray}
&& \hspace{-7mm} q_\mu \Gamma^{\mu \nu} = (p_f^2\!-\!p_i^2) P^\nu \theta_2 +[tP^\nu +  \tfrac{1}{2}(p_f^2\!-\!p_i^2) q^\nu] \theta_3 -  \tfrac{1}{2}q^\nu \theta_4. \nonumber \\ 
&&~ \hspace{-7mm} \label{eq:contr}
\end{eqnarray}
Since the four-vectors $P$ and $q$ are linearly independent, comparing their coefficients in Eqs.~(\ref{eq:WT2}) and (\ref{eq:contr}) we arrive at two relations:
\begin{eqnarray}
&& \hspace{-7mm} (p_f^2\!-\!p_i^2) \theta_2 + t  \theta_3 =  \Delta\!^{-1}(p_f^2) - \Delta\!^{-1}(p_i^2), \label{eq:rel1} \\
&& \hspace{-7mm} (p_f^2\!-\!p_i^2) \theta_3-\theta_4 =  -[\Delta\!^{-1}(p_f^2) + \Delta\!^{-1}(p_i^2)], \label{eq:rel2}
\end{eqnarray}
which is our key result.

Next, we carry out the procedure presented earlier for the charge form factors, now for the case of Eq.~ (\ref{eq:rel1}). 
At $t=0$ we have $(p_f^2\!-\!p_i^2) \theta_2(0,p_i^2,p_f^2) =  \Delta\!^{-1}(p_f^2) - \Delta\!^{-1}(p_i^2)$, therefore
\begin{eqnarray}
&& \hspace{-7mm} \theta_3(t,p_i^2,p_f^2) = \frac{(p_f^2\!-\!p_i^2)}{t} \left [  \theta_2(0,p_i^2,p_f^2) -  \theta_2(t,p_i^2,p_f^2)\right ], \nonumber \\
&&~\hspace{-7mm} \label{eq:t2ft3}
\end{eqnarray}
with $\theta_3(0,p_i^2,p_f^2) =(p_i^2\!-\!p_f^2)d\theta_2(t,p_i^2,p_f^2)/dt |_{t=0}$.
Moreover, comparing to Eq.~(\ref{eq:rulet0}), we find a relation between the off-shell gravitational and charge form factors at $t=0$,
\begin{eqnarray}
\theta_2(0,p_i^2,p_f^2)=F(0,p_i^2,p_f^2), \label{eq:F2}
\end{eqnarray} 
while for the half-off shell case
\begin{eqnarray}
\theta_2(0,m_\pi^2,p^2) = \theta_2(0,p^2,m_\pi^2) = \frac{\Delta\!^{-1}(p^2)}{(p^2-m_\pi^2)}, 
\end{eqnarray}
with $\theta_2(0,m_\pi^2,m_\pi^2)=1$ expressing the momentum sum
rule. From Eq.~(\ref{eq:rel2}) we find that
\begin{eqnarray}
&& \hspace{-7mm} \theta_4(t,p_i^2,p_f^2) = (p_f^2\!-\!p_i^2) \theta_3(t,p_i^2,p_f^2)  + \Delta\!^{-1}(p_f^2) + \Delta\!^{-1}(p_i^2). \nonumber \\ 
&& ~ \hspace{-7mm} \label{eq:rel3}
\end{eqnarray}
We note that $\theta_4$ vanishes if both the initial and final pions are on mass shell, $\theta_4(t,m_\pi^2,m_\pi^2)=0$, but in general is non-zero if either pion is off mass shell. The right-hand side can be expressed via $\theta_2$ only, 
\begin{eqnarray}
&& \hspace{-7mm} \theta_4(t,p_i^2,p_f^2) = \frac{(p_f^2\!-\!p_i^2)^2}{t} \left [  \theta_2(0,p_i^2,p_f^2) -  \theta_2(t,p_i^2,p_f^2)\right ]  \nonumber \\
&& \hspace{-6.7mm} \;+ (p_i^2-m_\pi^2) \theta_2(0,p_i^2,m_\pi^2) + (p_f^2-m_\pi^2) \theta_2(0,m_\pi^2,p_f^2).  \label{eq:rel4}
\end{eqnarray}
We remark that $\theta_4$ does not contribute to the moment in Eq.~(\ref{eq:poly}) upon the light-cone projection, 
as $n_\mu g^{\mu \nu} n_\nu = n^2=0$, where $n^\mu = (1,0,0,-1)/P^+$.
Note that $\theta_1$, which corresponds to a transverse tensor, does not enter into any constraints from the current conservation. In the chiral limit and the on-shell case of $m_\pi^2=0$ one has the low-energy theorem $\theta_1(0,0,0)= \theta_2(0,0,0)$~\cite{Donoghue:1991qv}.

In the last part of this Letter, we illustrate the above general
results in a quark model with spontaneously broken chiral symmetry,
treated at the one-loop (leading-$N_c$) level. One can
straightforwardly obtain expressions for the off-shell charge and
gravitational form factors in terms of the appropriate
Passarino-Veltman functions and explicitly verify that they comply to
all the relations provided above. The formulas become particularly
simple in the chiral limit in the Spectral Quark Model
(SQM)~\cite{RuizArriola:2003bs}, which is a one-loop chiral quark
model with the spectral function chosen in a way that enforces the vector
meson dominance of the pion charge form factor. The model is
consistent with the chiral, gauge and Lorentz invariance. In the chiral limit, $m_\pi=0$, manageably short
expressions emerge for the half-off-shell case:
\begin{eqnarray}
&& \hspace{-7mm} F(t,p^2,0)= \frac{M_V^4}{\left(M_V^2-p^2\right) \left(M_V^2-t\right)}, \\
&& \hspace{-7mm}  G(t,p^2,0)=\frac{p^2 M_V^2}{\left(M_V^2-p^2\right) \left(M_V^2-t\right)}, \nonumber \\
&& \hspace{-7mm} \theta_1(t,p^2,0)=\frac{M_V^2 \left[\frac{p^2 (t-p^2)}{M_V^2-p^2}+(t-2 p^2) L \right]}{\left(t-p^2\right)^2}, \nonumber \\
&& \hspace{-7mm} \theta_2(t,p^2,0)   = \frac{M_V^2 \left[\frac{p^2 (p^2-t)}{M_V^2-p^2}+t L\right]}{\left(t-p^2\right)^2}, \nonumber \\
&& \hspace{-7mm} \theta_3(t,p^2,0)  = \frac{p^2 M_V^2 \left[p^2-t+(M_V^2-p^2)L \right]}{\left(t-p^2\right)^2
   \left(M_V^2-p^2\right)}, \nonumber \\
&& \hspace{-7mm} \theta_4(t,p^2,0)  =   
   \frac{p^2 M_V^2 \left[\left(p^2\!-\!t\right) (2 p^2\!-\!t)+p^2 (M_V^2-p^2) L\right]}{(t-p^2)^2 (M_V^2-p^2)}, \nonumber 
\end{eqnarray}
with $L=\log\frac{M_V^2-p^2}{M_V^2-t}$ and $M_V$ denoting the $\rho$
meson mass. We note that while in this model, where the inverse pion propagator 
is $\Delta\!^{-1}(p^2)= {M_V^2 p^2 }/(M_V^2-p^2)$, the charged form factors
exhibit a factorized form, this is not the case of the gravitational
form factors. 
The above formulas satisfy all the general relations above, namely
Eqs.~(\ref{eq:Grel}, \ref{eq:t2ft3}, \ref{eq:F2}, \ref{eq:rel3})
and (\ref{eq:rel4}). It is thus tempting to make a first estimate of the 
off-shell effects in the pion GPDs in a model implementing these new 
constraints. 

\begin{figure}[t]
\centering
\includegraphics[angle=0,width=0.43 \textwidth]{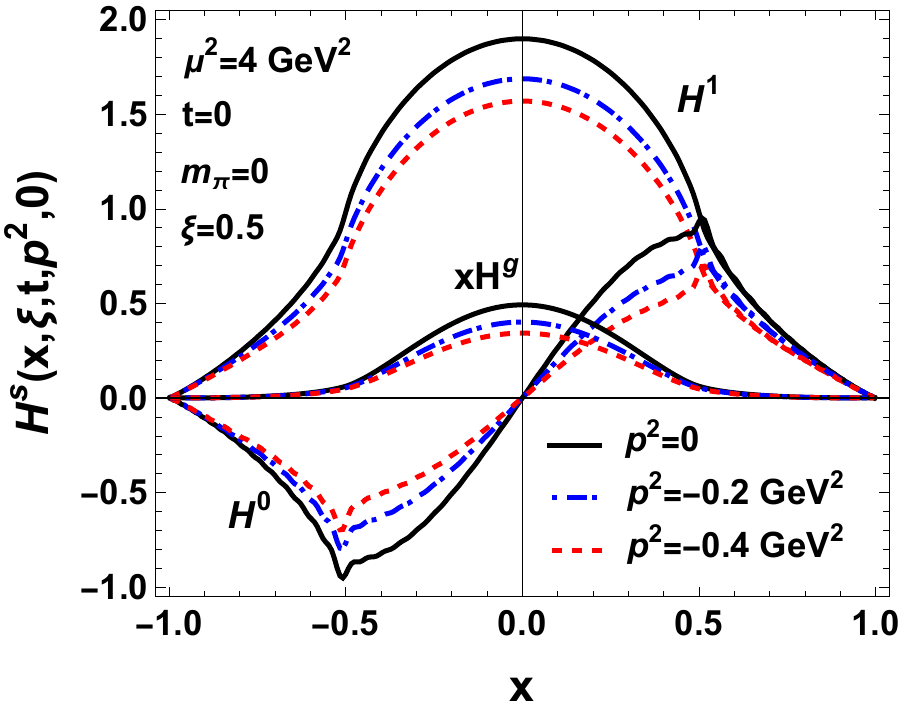} 
\vspace{-1mm}
\caption{Half-off-shell effective GPDs of the pion obtained in SQM in the chiral limit for $\xi=0.5$, $t=0$, and evolved to $\mu^2=4~{\rm GeV}^2$. Line types distinguish the off-shellness, whereas the three bundles correspond to the quark isovector GPD, quark isoscalar GPD, and the gluon GPD multiplied with $x$. \label{fig:gpd}}
\end{figure}

The on-shell GPDs are obtained in SQM at the quark model
scale~\cite{Broniowski:2007si}, $\mu_0$, where the valence quarks
carry $100\%$ of the momentum, and are subsequently evolved to a
higher scale $\mu$ with the leading-order DGLAP-ERBL equations
~\cite{GolecBiernat:1998ja}. The half-off-shell GPDs at $\mu_0$
involve rather lengthy analytic formulas (not shown here for brevity)
and display a lack of factorization in $x$, $t$, or $p^2$.  A sample
result (with $t=0$ and $\xi=0.5$) at $\mu = 2$~GeV is presented in
Fig.~\ref{fig:gpd}. We note a significant dependence on the
(space-like) off-shell parameter $p^2$. For the difference at the
maxima of the curves at $p^2=-0.2~{\rm GeV}^2$ and $p^2=0$ we note the
relative effect of 10\% for the isovector GPD, and somewhat larger
18\% for the isoscalar GPDs. For $p^2=-0.4~{\rm GeV}^2$ the effects,
are, correspondingly, 20\% and 35\%. As expected, the size depends on
$p^2$, which is controlled by the kinematics of the Sullivan
process. The effect reflects qualitatively the change of
normalization with $p^2$ according to Eq.~(\ref{eq:poly2}). This
feature becomes exact at $\mu\to\infty$, as then the GPDs become
localized in the ERBL region $|x|\le\xi$ (see
eg.~\cite{Broniowski:2007si}), with the normalization given by
Eq.~(\ref{eq:poly2}), and the relative normalization of $xH^0$ and
$xH^g$ given by the ratio $3N_f/16$, where $N_f=3$ is the number of
active flavors.

The sources of model uncertainty to absolute (not relative) values of
the GPDs include the value of the vector meson mass in SQM (which
attributes a roughly 10-15\% effect to form
factors~\cite{Masjuan:2012sk} and to parton distributions), the
approximation of the exact chiral limit (about
5\%~\cite{Davidson:1994uv}), and the uncertainty in the value of the
quark-model scale (about 10\%~\cite{Broniowski:2007si}). 

Finally, we wish to digress on a relevant methodological point, along
the lines of~\cite{Koch:2001ii}, concerning evaluation of amplitudes
such as in Fig.~\ref{fig:sul}. The off-shellness affects, in general,
all  the components of the diagram. In our case, it influences the GPD (as
discussed above), hence the DVCS amplitude, but also the pion form
factor (as well as the pion nucleon form factor, not discussed
here). The leg amputation procedure of Eq.~(\ref{eq:leg}) presumes
that in the propagator attached to the vertex is the full pion
propagator, with off-shell effects, and not its pole
approximation. Hence we should just use $\Delta(p^2)$ to maintain
consistency.  If however, as is typically done phenomenologically, one
admitted the pion pole term only, $1/(p^2-m_\pi^2)$, one would miss
the factor $\Delta(p^2) (p^2-m_\pi^2)=1/F(0,p^2,m_\pi^2)$. This factor
could be conventionally attributed to the half-off-shell vertex, by
introducing
$\Gamma^{\ast \mu}(t,p^2,m_\pi^2)\equiv \Gamma^\mu(t,p^2,m_\pi^2)/F(0,p^2,m_\pi^2)$,
to be used in calculations with the attached pion propagators taken as
a pole term. In our case, for the charge form factor we have
\begin{eqnarray}
\Gamma^{\ast \mu}(t,p^2,m_\pi^2)&=&2P^\mu \frac{F(t,p^2,m_\pi^2)}{F(0,p^2,m_\pi^2)}\label{eq:star} \\
&-&  q^\mu \frac{p^2}{t} \left[ 1 -\frac{F(t,p^2,m_\pi^2)}{F(0,p^2,m_\pi^2)} \right]. \nonumber
\end{eqnarray}
If the dependence on $t$ and $p^2$ in $F(t,p^2,m_\pi^2)$ factorizes
(as is the case of SQM in the chiral limit but not in general), then
the only dependence on $p^2$ sits in front of the factor associated
with the $q^\mu$ part, which is always present for virtual photons
(for a real photon it may be removed by the choice of
gauge~\cite{Koch:2001ii}). This would result in a $p^2$ independent
form factor $G/p^2$, as considered in~\cite{Choi:2019nvk}.  Similarly
to Eq.~(\ref{eq:star}), one could attribute the propagator correction
$1/F(0,p^2,m_\pi^2)$ to the half-off-shell GPDs.

Finally, we remind that as discussed in~\cite{Amrath:2008vx}, the
deeply virtual Compton scattering (DVCS) amplitude, involving  the
GPDs, enters the cross section formula for the Sullivan process via
interference with the Bethe-Heitler amplitude, thus uncertainties in
the GPDs carry over linearly to the cross section.  That way, the off-shellness
contributes to the uncertainties encountered in the modeling of the
Sullivan process, together with such quantities as the parton
distributions (taken on shell), pion propagator (possibility of
inclusion of the excited states) or the pion-nucleon form factor.

To summarize, we have considered, on a general footing, the off-shell
GPDs of the pion and the related electromagnetic and gravitational
form factors. We have shown that the WTI for the energy-stress tensor
results in relations between the four off-shell gravitational form
factors, in analogy to the case of the two off-shell electromagnetic
form factors. These relations may serve as consistency constraints in
constructing phenomenological off-shell GPDs.  We have employed a
simple chiral quark model to illustrate the general formalism, as well
as to assess the actual size of the effects after the QCD evolution to
the scale $\mu=2$~GeV. We find a non-negligible (roughly, 10\%)
influence already at off-shellness of the order of $-0.2~{\rm GeV}^2$,
especially when $\xi$ is not close to 0. We finally note that our analysis can
be straightforwardly extended to the other members of the pseudoscalar
nonet, in particular the kaons, for which the Sullivan process at the
EIC is also currently being considered~\cite{Arrington:2021biu}.

\medskip

We are grateful to Krzysztof Golec-Biernat for providing us with his QCD evolution code.
VS acknowledges the support by the Polish National Science Centre (NCN), grant 2019/33/B/ST2/00613, WB by NCN, grant 2018/31/B/ST2/01022, and ERA by project PID2020-114767GB-I00 funded by MCIN/AEI/10.13039/501100011033 as well as Junta de Andaluc{\'i}a (grant FQM-225).
 
\bibliography{Ref}

\begin{thebibliography}{24}
\expandafter\ifx\csname natexlab\endcsname\relax\def\natexlab#1{#1}\fi
\providecommand{\url}[1]{\texttt{#1}}
\providecommand{\href}[2]{#2}
\providecommand{\path}[1]{#1}
\providecommand{\DOIprefix}{doi:}
\providecommand{\ArXivprefix}{arXiv:}
\providecommand{\URLprefix}{URL: }
\providecommand{\Pubmedprefix}{pmid:}
\providecommand{\doi}[1]{\href{http://dx.doi.org/#1}{\path{#1}}}
\providecommand{\Pubmed}[1]{\href{pmid:#1}{\path{#1}}}
\providecommand{\bibinfo}[2]{#2}
\ifx\xfnm\relax \def\xfnm[#1]{\unskip,\space#1}\fi
\bibitem[{Aguilar et~al.(2019)}]{Aguilar:2019teb}
\bibinfo{author}{A.~C. Aguilar}, et~al.,
\newblock \bibinfo{title}{{Pion and kaon structure at the Electron-Ion
  Collider}},
\newblock \bibinfo{journal}{Eur. Phys. J. A} \bibinfo{volume}{55}
  (\bibinfo{year}{2019}) \bibinfo{pages}{190}.
  \DOIprefix\doi{10.1140/epja/i2019-12885-0}.
  \href{http://arxiv.org/abs/1907.08218}{{\tt arXiv:1907.08218}}.
\bibitem[{Ch\'avez et~al.(2022)Ch\'avez, Bertone, De~Soto~Borrero, Defurne,
  Mezrag, Moutarde, Rodr\'\i{}guez-Quintero, and Segovia}]{Chavez:2021koz}
\bibinfo{author}{J.~M.~M. Ch\'avez}, \bibinfo{author}{V.~Bertone},
  \bibinfo{author}{F.~De~Soto~Borrero}, \bibinfo{author}{M.~Defurne},
  \bibinfo{author}{C.~Mezrag}, \bibinfo{author}{H.~Moutarde},
  \bibinfo{author}{J.~Rodr\'\i{}guez-Quintero}, \bibinfo{author}{J.~Segovia},
\newblock \bibinfo{title}{{Accessing the Pion 3D Structure at US and China
  Electron-Ion Colliders}},
\newblock \bibinfo{journal}{Phys. Rev. Lett.} \bibinfo{volume}{128}
  (\bibinfo{year}{2022}) \bibinfo{pages}{202501}.
  \DOIprefix\doi{10.1103/PhysRevLett.128.202501}.
  \href{http://arxiv.org/abs/2110.09462}{{\tt arXiv:2110.09462}}.
\bibitem[{Ji(1997)}]{Ji:1996nm}
\bibinfo{author}{X.-D. Ji},
\newblock \bibinfo{title}{{Deeply virtual Compton scattering}},
\newblock \bibinfo{journal}{Phys. Rev. D} \bibinfo{volume}{55}
  (\bibinfo{year}{1997}) \bibinfo{pages}{7114--7125}.
  \DOIprefix\doi{10.1103/PhysRevD.55.7114}.
  \href{http://arxiv.org/abs/hep-ph/9609381}{{\tt arXiv:hep-ph/9609381}}.
\bibitem[{Radyushkin(1997)}]{Radyushkin:1997ki}
\bibinfo{author}{A.~V. Radyushkin},
\newblock \bibinfo{title}{{Nonforward parton distributions}},
\newblock \bibinfo{journal}{Phys. Rev. D} \bibinfo{volume}{56}
  (\bibinfo{year}{1997}) \bibinfo{pages}{5524--5557}.
  \DOIprefix\doi{10.1103/PhysRevD.56.5524}.
  \href{http://arxiv.org/abs/hep-ph/9704207}{{\tt arXiv:hep-ph/9704207}}.
\bibitem[{Sullivan(1972)}]{PhysRevD.5.1732}
\bibinfo{author}{J.~D. Sullivan},
\newblock \bibinfo{title}{One-pion exchange and deep-inelastic electron-nucleon
  scattering},
\newblock \bibinfo{journal}{Phys. Rev. D} \bibinfo{volume}{5}
  (\bibinfo{year}{1972}) \bibinfo{pages}{1732--1737}.
  \DOIprefix\doi{10.1103/PhysRevD.5.1732}.
\bibitem[{Burkardt(2000)}]{Burkardt:2000za}
\bibinfo{author}{M.~Burkardt},
\newblock \bibinfo{title}{{Impact parameter dependent parton distributions and
  off forward parton distributions for $\zeta \to 0$}},
\newblock \bibinfo{journal}{Phys. Rev. D} \bibinfo{volume}{62}
  (\bibinfo{year}{2000}) \bibinfo{pages}{071503}.
  \DOIprefix\doi{10.1103/PhysRevD.62.071503}.
  \href{http://arxiv.org/abs/hep-ph/0005108}{{\tt arXiv:hep-ph/0005108}},
  \bibinfo{note}{[Erratum: Phys.Rev.D 66, 119903 (2002)]}.
\bibitem[{Diehl(2003)}]{Diehl:2003ny}
\bibinfo{author}{M.~Diehl},
\newblock \bibinfo{title}{Generalized parton distributions},
\newblock \bibinfo{journal}{Phys. Rept.} \bibinfo{volume}{388}
  (\bibinfo{year}{2003}) \bibinfo{pages}{41--277}.
  \DOIprefix\doi{10.1016/j.physrep.2003.08.002}.
  \href{http://arxiv.org/abs/hep-ph/0307382}{{\tt arXiv:hep-ph/0307382}}.
\bibitem[{M\"uller et~al.(1994)M\"uller, Robaschik, Geyer, Dittes, and
  Ho\v{r}ej\v{s}i}]{Muller:1994ses}
\bibinfo{author}{D.~M\"uller}, \bibinfo{author}{D.~Robaschik},
  \bibinfo{author}{B.~Geyer}, \bibinfo{author}{F.~M. Dittes},
  \bibinfo{author}{J.~Ho\v{r}ej\v{s}i},
\newblock \bibinfo{title}{{Wave functions, evolution equations and evolution
  kernels from light ray operators of QCD}},
\newblock \bibinfo{journal}{Fortsch. Phys.} \bibinfo{volume}{42}
  (\bibinfo{year}{1994}) \bibinfo{pages}{101--141}.
  \DOIprefix\doi{10.1002/prop.2190420202}.
  \href{http://arxiv.org/abs/hep-ph/9812448}{{\tt arXiv:hep-ph/9812448}}.
\bibitem[{Nishijima and Singh(1967)}]{Nishijima:1967byg}
\bibinfo{author}{K.~Nishijima}, \bibinfo{author}{A.~H. Singh},
\newblock \bibinfo{title}{{Normalization of Bethe-Salpeter amplitudes}},
\newblock \bibinfo{journal}{Phys. Rev.} \bibinfo{volume}{162}
  (\bibinfo{year}{1967}) \bibinfo{pages}{1740}.
  \DOIprefix\doi{10.1103/PhysRev.162.1740}.
\bibitem[{Naus and Koch(1989)}]{Naus:1989em}
\bibinfo{author}{H.~W.~L. Naus}, \bibinfo{author}{J.~H. Koch},
\newblock \bibinfo{title}{{Use of form-factors in electromagnetic
  interactions}},
\newblock \bibinfo{journal}{Phys. Rev. C} \bibinfo{volume}{39}
  (\bibinfo{year}{1989}) \bibinfo{pages}{1907--1913}.
  \DOIprefix\doi{10.1103/PhysRevC.39.1907}.
\bibitem[{Rudy et~al.(1994)Rudy, Fearing, and Scherer}]{Rudy:1994qb}
\bibinfo{author}{T.~E. Rudy}, \bibinfo{author}{H.~W. Fearing},
  \bibinfo{author}{S.~Scherer},
\newblock \bibinfo{title}{{The off-shell electromagnetic form-factors of pions
  and kaons in chiral perturbation theory}},
\newblock \bibinfo{journal}{Phys. Rev. C} \bibinfo{volume}{50}
  (\bibinfo{year}{1994}) \bibinfo{pages}{447--459}.
  \DOIprefix\doi{10.1103/PhysRevC.50.447}.
  \href{http://arxiv.org/abs/hep-ph/9401302}{{\tt arXiv:hep-ph/9401302}}.
\bibitem[{Schnitzer and Weinberg(1967)}]{Schnitzer:1967zzb}
\bibinfo{author}{H.~J. Schnitzer}, \bibinfo{author}{S.~Weinberg},
\newblock \bibinfo{title}{{Current-algebra calculation of hard-pion processes:
  $A_1 \to \rho+\pi$ and $\rho \to \pi+\pi$}},
\newblock \bibinfo{journal}{Phys. Rev.} \bibinfo{volume}{164}
  (\bibinfo{year}{1967}) \bibinfo{pages}{1828--1833}.
  \DOIprefix\doi{10.1103/PhysRev.164.1828}.
\bibitem[{Choi et~al.(2019)Choi, Frederico, Ji, and de~Melo}]{Choi:2019nvk}
\bibinfo{author}{H.-M. Choi}, \bibinfo{author}{T.~Frederico},
  \bibinfo{author}{C.-R. Ji}, \bibinfo{author}{J.~P. B.~C. de~Melo},
\newblock \bibinfo{title}{{Pion off-shell electromagnetic form factors: data
  extraction and model analysis}},
\newblock \bibinfo{journal}{Phys. Rev. D} \bibinfo{volume}{100}
  (\bibinfo{year}{2019}) \bibinfo{pages}{116020}.
  \DOIprefix\doi{10.1103/PhysRevD.100.116020}.
  \href{http://arxiv.org/abs/1908.01185}{{\tt arXiv:1908.01185}}.
\bibitem[{Raman(1971)}]{Raman:1971jg}
\bibinfo{author}{K.~Raman},
\newblock \bibinfo{title}{{Gravitational form-factors of pseudoscalar mesons,
  stress-tensor-current commutation relations, and deviations from tensor- and
  scalar-meson dominance}},
\newblock \bibinfo{journal}{Phys. Rev. D} \bibinfo{volume}{4}
  (\bibinfo{year}{1971}) \bibinfo{pages}{476--488}.
  \DOIprefix\doi{10.1103/PhysRevD.4.476}.
\bibitem[{Brout and Englert(1966)}]{Brout:1966oea}
\bibinfo{author}{R.~Brout}, \bibinfo{author}{F.~Englert},
\newblock \bibinfo{title}{{Gravitational Ward identity and the principle of
  equivalence}},
\newblock \bibinfo{journal}{Phys. Rev.} \bibinfo{volume}{141}
  (\bibinfo{year}{1966}) \bibinfo{pages}{1231--1232}.
  \DOIprefix\doi{10.1103/PhysRev.141.1231}.
\bibitem[{Donoghue and Leutwyler(1991)}]{Donoghue:1991qv}
\bibinfo{author}{J.~F. Donoghue}, \bibinfo{author}{H.~Leutwyler},
\newblock \bibinfo{title}{{Energy and momentum in chiral theories}},
\newblock \bibinfo{journal}{Z. Phys. C} \bibinfo{volume}{52}
  (\bibinfo{year}{1991}) \bibinfo{pages}{343--351}.
  \DOIprefix\doi{10.1007/BF01560453}.
\bibitem[{Ruiz~Arriola and Broniowski(2003)}]{RuizArriola:2003bs}
\bibinfo{author}{E.~Ruiz~Arriola}, \bibinfo{author}{W.~Broniowski},
\newblock \bibinfo{title}{{Spectral quark model and low-energy hadron
  phenomenology}},
\newblock \bibinfo{journal}{Phys. Rev.} \bibinfo{volume}{D67}
  (\bibinfo{year}{2003}) \bibinfo{pages}{074021}.
  \DOIprefix\doi{10.1103/PhysRevD.67.074021}.
  \href{http://arxiv.org/abs/hep-ph/0301202}{{\tt arXiv:hep-ph/0301202}}.
\bibitem[{Broniowski et~al.(2008)Broniowski, Arriola, and
  Golec-Biernat}]{Broniowski:2007si}
\bibinfo{author}{W.~Broniowski}, \bibinfo{author}{E.~R. Arriola},
  \bibinfo{author}{K.~Golec-Biernat},
\newblock \bibinfo{title}{{Generalized parton distributions of the pion in
  chiral quark models and their QCD evolution}},
\newblock \bibinfo{journal}{Phys. Rev. D} \bibinfo{volume}{77}
  (\bibinfo{year}{2008}) \bibinfo{pages}{034023}.
  \DOIprefix\doi{10.1103/PhysRevD.77.034023}.
  \href{http://arxiv.org/abs/0712.1012}{{\tt arXiv:0712.1012}}.
\bibitem[{Golec-Biernat and Martin(1999)}]{GolecBiernat:1998ja}
\bibinfo{author}{K.~J. Golec-Biernat}, \bibinfo{author}{A.~D. Martin},
\newblock \bibinfo{title}{Off-diagonal parton distributions and their
  evolution},
\newblock \bibinfo{journal}{Phys. Rev.} \bibinfo{volume}{D59}
  (\bibinfo{year}{1999}) \bibinfo{pages}{014029}.
  \href{http://arxiv.org/abs/hep-ph/9807497}{{\tt arXiv:hep-ph/9807497}}.
\bibitem[{Masjuan et~al.(2013)Masjuan, Ruiz~Arriola, and
  Broniowski}]{Masjuan:2012sk}
\bibinfo{author}{P.~Masjuan}, \bibinfo{author}{E.~Ruiz~Arriola},
  \bibinfo{author}{W.~Broniowski},
\newblock \bibinfo{title}{{Meson dominance of hadron form factors and
  large-$N_c$ phenomenology}},
\newblock \bibinfo{journal}{Phys. Rev. D} \bibinfo{volume}{87}
  (\bibinfo{year}{2013}) \bibinfo{pages}{014005}.
  \DOIprefix\doi{10.1103/PhysRevD.87.014005}.
  \href{http://arxiv.org/abs/1210.0760}{{\tt arXiv:1210.0760}}.
\bibitem[{Davidson and Ruiz~Arriola(1995)}]{Davidson:1994uv}
\bibinfo{author}{R.~Davidson}, \bibinfo{author}{E.~Ruiz~Arriola},
\newblock \bibinfo{title}{{Structure functions of pseudoscalar mesons in the
  SU(3) NJL model}},
\newblock \bibinfo{journal}{Phys.Lett.} \bibinfo{volume}{B348}
  (\bibinfo{year}{1995}) \bibinfo{pages}{163--169}.
  \DOIprefix\doi{10.1016/0370-2693(95)00091-X}.
\bibitem[{Koch et~al.(2002)Koch, Pascalutsa, and Scherer}]{Koch:2001ii}
\bibinfo{author}{J.~H. Koch}, \bibinfo{author}{V.~Pascalutsa},
  \bibinfo{author}{S.~Scherer},
\newblock \bibinfo{title}{{Hadron structure and the limitations of
  phenomenological models in electromagnetic reactions}},
\newblock \bibinfo{journal}{Phys. Rev. C} \bibinfo{volume}{65}
  (\bibinfo{year}{2002}) \bibinfo{pages}{045202}.
  \DOIprefix\doi{10.1103/PhysRevC.65.045202}.
  \href{http://arxiv.org/abs/nucl-th/0108044}{{\tt arXiv:nucl-th/0108044}}.
\bibitem[{Amrath et~al.(2008)Amrath, Diehl, and Lansberg}]{Amrath:2008vx}
\bibinfo{author}{D.~Amrath}, \bibinfo{author}{M.~Diehl}, \bibinfo{author}{J.-P.
  Lansberg},
\newblock \bibinfo{title}{{Deeply virtual Compton scattering on a virtual pion
  target}},
\newblock \bibinfo{journal}{Eur. Phys. J. C} \bibinfo{volume}{58}
  (\bibinfo{year}{2008}) \bibinfo{pages}{179--192}.
  \DOIprefix\doi{10.1140/epjc/s10052-008-0769-1}.
  \href{http://arxiv.org/abs/0807.4474}{{\tt arXiv:0807.4474}}.
\bibitem[{Arrington et~al.(2021)}]{Arrington:2021biu}
\bibinfo{author}{J.~Arrington}, et~al.,
\newblock \bibinfo{title}{{Revealing the structure of light pseudoscalar mesons
  at the electron\textendash{}ion collider}},
\newblock \bibinfo{journal}{J. Phys. G} \bibinfo{volume}{48}
  (\bibinfo{year}{2021}) \bibinfo{pages}{075106}.
  \DOIprefix\doi{10.1088/1361-6471/abf5c3}.
  \href{http://arxiv.org/abs/2102.11788}{{\tt arXiv:2102.11788}}.

\end{thebibliography}

\end{document}